\documentclass[%
a4paper,       
UKenglish,     
DIV10,          
useAMS,usenatbib,referee, doublespace]%
{scrartcl}\usepackage[]{graphicx}\usepackage[]{color}
\makeatletter
\def\maxwidth{ %
  \ifdim\Gin@nat@width>\linewidth
    \linewidth
  \else
    \Gin@nat@width
  \fi
}
\makeatother

\definecolor{fgcolor}{rgb}{0.345, 0.345, 0.345}

\usepackage{framed}
\makeatletter
 {\par\unskip\endMakeFramed%
 \at@end@of@kframe}
\makeatother

\definecolor{shadecolor}{rgb}{.97, .97, .97}
\definecolor{messagecolor}{rgb}{0, 0, 0}
\definecolor{warningcolor}{rgb}{1, 0, 1}
\definecolor{errorcolor}{rgb}{1, 0, 0}
\newenvironment{knitrout}{}{} 

\usepackage{alltt}

\usepackage[utf8]{inputenc}
\usepackage[UKenglish]{babel}
\usepackage[T1]{fontenc}
\usepackage{textcomp}
\usepackage{setspace}
\usepackage{hyperref}  
\usepackage{url}                
\usepackage[sumlimits, intlimits, namelimits]{amsmath} 
\usepackage{amssymb}            
\usepackage[caption = true]{subfig} 
\usepackage{array}              
\usepackage{booktabs}           
\usepackage{ae}                 
\usepackage[longnamesfirst,round]{natbib} 
\usepackage{color}              
\usepackage[sc]{mathpazo}       
\usepackage{helvet}             
\usepackage{todonotes}          

\newcommand{\hl}{\textcolor{black}}

\hypersetup{
  bookmarksopen=true, 
  breaklinks=true,
pdftitle={A New Standard for the Analysis and Design of Replication Studies}, 
  pdfsubject={Statistics},
  pdfkeywords={Statistics, Evidence},
  colorlinks=true,
  linkcolor=blue,
  anchorcolor=black,
  citecolor=teal,
  urlcolor=purple
}

\usepackage{xifthen}            
\usepackage{array}
\usepackage{amssymb}

\newcommand{\blanco}[1]{  } 

\newcommand{\latin}[1]{\textit{#1}}

\newcommand{\abk}[1]{\mbox{#1}\xdot}
\DeclareRobustCommand\xdot{\futurelet\token\Xdot}
\def\Xdot{%
  \ifx\token\bgroup.%
  \else\ifx\token\egroup.%
  \else\ifx\token\/.%
  \else\ifx\token\ .%
  \else\ifx\token!.%
  \else\ifx\token,.%
  \else\ifx\token:.%
  \else\ifx\token;.%
  \else\ifx\token?.%
  \else\ifx\token/.%
  \else\ifx\token'.%
  \else\ifx\token).%
  \else\ifx\token-.%
  \else\ifx\token+.%
  \else\ifx\token~.%
  \else\ifx\token.%
  \else.\ %
  \fi\fi\fi\fi\fi\fi\fi\fi\fi\fi\fi\fi\fi\fi\fi\fi%
}

\newcommand{\eg}{\abk{\latin{e.\,g}}}
\newcommand{\ie}{\abk{\latin{i.\,e}}}

\newlength{\halbebreite}
\DeclareMathOperator{\Nor}{N} 
  
\DeclareMathOperator{\Be}{Be} 

\DeclareMathOperator{\se}{se}   
\DeclareMathOperator{\Med}{Med} 

\newcommand{\partialv}[3][1]{%
\ifthenelse{#1 = 1}{\frac{\partial #2}{\partial #3}}{\frac{\partial^{#1} #2}{\partial #3^{#1}}}
} 

\newcommand{\partials}[3][1]{%
\ifthenelse{#1 = 1}{\frac{d #2}{d #3}}{\frac{d^{#1} #2}{d #3^{#1}}}
} 

\newcommand{\dseps}[2][1]{%
\ifthenelse{#1 = 1}{\frac{d}{d #2}}{\frac{d^{#1}}{d #2^{#1}}}
}

\newcommand{\dsepv}[2][1]{%
\ifthenelse{#1 = 1}{\frac{\partial}{\partial #2}}{\frac{\partial^{#1}}{\partial #2^{#1}}}
}

\newcommand{\ml}[2][1]{
\ifthenelse{#1 = 1}%
 {\hat{#2}_{\scriptscriptstyle{\mathrm{ML}}}}%
 {\hat{#2}^{#1}_{\scriptscriptstyle{\mathrm{ML}}}}
}
\newcommand{\map}[2][0]{
\ifthenelse{#1 = 0}%
 {\hat{#2}_{\scriptscriptstyle{\mathrm{MAP}}}}%
 {\hat{#2}_{{\scriptscriptstyle{\mathrm{MAP}}_{#1}}}}
}
\newcommand{\mpm}[2][0]{
\ifthenelse{#1 = 0}%
 {\hat{#2}_{\scriptscriptstyle{\mathrm{MPM}}}}%
 {\hat{#2}_{{\scriptscriptstyle{\mathrm{MPM}}_{#1}}}}
}

\newcommand{\given}{\,\vert\,} 

\newcommand{\abs}[1]{\left\lvert#1\right\rvert} 

\IfFileExists{upquote.sty}{\usepackage{upquote}}{}
\IfFileExists{upquote.sty}{\usepackage{upquote}}{}
\begin{document}

\title{A New Standard for the Analysis and Design of Replication Studies}
\author{Leonhard
  Held\\ 
  Epidemiology, Biostatistics
  and Prevention Institute (EBPI)\\ Center for Reproducible Science  \\ University of Zurich\\ Hirschengraben 84,
  8001 Zurich, Switzerland\\ Email: \texttt{leonhard.held@uzh.ch}}

\maketitle

\begin{center}
\begin{minipage}{12cm}
  \textbf{Abstract}: A new standard is proposed for the evidential
  assessment of replication studies.  The approach combines a specific
  reverse-Bayes technique with prior-predictive tail
  probabilities to define replication success.  The method gives rise
  to a quantitative measure for replication success, called the
  sceptical $p$-value.  The sceptical $p$-value inte\-grates traditional
  significance of both the original and replication study with a
  comparison of the respective effect sizes.  It incorporates the
  uncertainty of both the original and replication effect estimates
  and reduces to the ordinary $p$-value of the replication study if
  the uncertainty of the original effect estimate is ignored.  The
  proposed framework can also be used to determine the power or the
  required \hl{replication} sample size to achieve replication success.  Numerical
  calculations highlight the difficulty to achieve replication success if
  the evidence from the original study is only suggestive.  An
  application to data from the Open Science Collaboration project on
  the replicability of
  psychological science illustrates the proposed methodology.\\
\noindent
  \textbf{Key Words}: {Power; \hl{Prior-Data Conflict}; Replication Success; \hl{Reverse-Bayes}; Sample Size; Sceptical $p$-value}
\end{minipage}
\end{center}

\section{Introduction}
Replicability of research findings is crucial to the credibility of
all empirical domains of science.  As a consequence of the so-called
replication crisis \citep{Ioannidis2005,Begley2015}, the past years
have witnessed increasing interest in large-scale
replication projects, \eg
\cite{OSC2015,Camerer2016,Camerer2018}. Such efforts help 
assess to what extent claims of new discoveries can be confirmed in
independent replication studies whose procedures are as closely matched to the
original studies as possible.

However, there is no established standard for the statistical evaluation of
replication success.  Standard significance of the
replication study is often used \hl{as a criterion, but significance alone does not take the effect sizes of the original and replication study into account and}
can easily lead to conclusions opposite
to what the evidence warrants \citep{Simonsohn2015}. A comparison of
the effect sizes of the original and replication study is also common,
where a smaller replication effect estimate decreases the credibility
of the original study result. A modification of this is to investigate
whether the replication effect estimate is compatible with the
original effect estimate \citep{BayarriMayoral2002,Patil2016}. 
\hl{Meta-analytic approaches take
the results from the original and replication study at face value and
combine them into an overall effect size estimate.  However, when
conducting a replication, researchers are challenging the original
study and asking {\em whether} they should take it at face value.  This is
an inherently asymmetric question, where exchangeability assumptions
are not appropriate and alternative methods for evidence
quantification are
needed. }

Recently the lack of a single accepted definition of replicability has
been emphasized by \cite{Goodman2016} who call for a better
understanding of the relationship between reproducibility and the
truth of scientific claims.  Researchers have started to develop Bayesian methods to analyse and design replication studies
\citep{VerhagenWagenmakers2014,vanAertvanAssen2017,Schoenbrodt2018}, but there is a lack
of appropriate methodology based on traditional metrics (effect
estimates, confidence intervals and $p$-values). To address this
deficiency, I propose a principled approach, combining the analysis of
credibility \citep{matthews:2001,matthews:2001b} with the prior criticism approach 
by \cite{box:1980} and \cite{evans.moshonov2006} to define {\em replication
  success} (Section \ref{sec:AnCred}). This gives rise to a new
quantitative measure of replication success, the {\em sceptical
  $p$-value} (Section \ref{sec:sceptical.p}).

The sceptical $p$-value has attractive properties. It takes into
account the results from both the original and the replication
study and is always larger than the ordinary
$p$-values from these studies. If 
the uncertainty of the original effect estimate is ignored, the sceptical
$p$-value reduces to the ordinary $p$-value from the replication
study. 
Moreover, the sceptical $p$-value considers replication studies
with relatively small effect estimates (compared to the original estimates) as less successful. 
To avoid the so-called replication paradox
\citep{Ly_etal2018}, a one-sided sceptical $p$-value is derived
within the proposed framework, ensuring that replication success can
only occur if the original and replication effect estimates have the
same sign.

Statistical power is of central importance in assessing the
reliability of science \citep{Button_etal2013}. Appropriate design of
a replication study is key to tackling the replication crisis as many
such studies are currently severely under-powered, even by traditional
standards \citep{Anderson2017}.  
Methods to calculate the power for replication success are proposed
in Section \ref{sec:design}. Numerical calculations highlight the
difficulty to achieve replication success if the evidence from the
original study is only suggestive.  The framework is also 
used to determine the required sample size to achieve replication
success with appropriate power.  Section \ref{sec:sec5} presents a
reanalysis of data from the  \citet{OSC2015} project on the replicability of psychological science 
to illustrate
the usefulness of the proposed methodology. I close with some comments
in Section \ref{sec:sec6}.

\section{Assessment of Replication Success}
\label{sec:AnCred}
Analysis of credibility \citep{matthews:2001,matthews:2001b} is a
reverse-Bayes procedure originally designed to assess the credibility
of significant findings in the light of existing evidence.  \hl{The
  discussion of \citet{matthews:2001b} and the author's response
  provide additional insights on the philosophy and detail of this
  method.} The idea to use Bayes's theorem in reverse originates in
the work of I.J.~Good \citep{good:1950,good:1983} and is increasingly
used to assess the plausibility of scientific findings
\citep{greenland:2006,greenland:2011,held2013,Colquhoun:2017,Colquhoun:2019}. 

\subsection{\hl{Reverse-Bayes Analysis}}
Analysis of credibility combines a significant effect estimate
from the original study with a sceptical prior
\cite[Section 4.1.3]{Spiegelhalter1994}\hl{, a normal distribution centered} around zero to
represent doubts about large effect estimates.  A sceptical prior
shrinks the original effect estimate towards zero, where the amount of
shrinkage depends on the sceptical prior variance.
\cite{Fletcher1993} have argued for the use of sceptical priors for
original clinical study results, which often show a tendency for
overoptimism.

In order to challenge the original study it is natural to ask how
sceptical we would have to be not to find its apparently positive
effect estimate convincing. This leads to a reverse-Bayes approach,
where the posterior is fixed to have a lower (or upper) credible
limit exactly equal to zero and the sceptical prior variance is chosen
accordingly.  The approach thus represents the objection by a sceptic
who argues that the original result would no longer be 'significant'
if combined with a sufficiently sceptical prior. The goal is now to
persuade the sceptic by showing that this prior is unrealistic. To do
so, a replication study is conducted.  If the data from the
replication study are in conflict with the sufficiently sceptical
prior, the original study result is confirmed.

Suppose the original study gives rise to a
conventional confidence interval for the unknown effect size $\theta$
at level $1-\alpha$ with lower limit $L$ and upper limit $U$.  
Assume that $L$ and $U$ are symmetric around the
original point estimate $\hat \theta_o$ (assumed to be normally
distributed) and that both are either positive or negative, \ie the
original effect is significant at significance level $\alpha$.  After
a suitable transformation this framework covers a large number of
commonly used effect measures such as differences in means, odds
ratios, relative risks and correlations.

We first need to compute the variance of the sufficiently sceptical
prior.  \citet{matthews:2001} has shown that the equi-tailed credible
interval of the sufficiently sceptical prior at level $1-\alpha$ has
limits $\pm S$ where
\begin{equation}\label{eq:S}
S = \frac{(U-L)^2}{4 \sqrt{UL}}
\end{equation}
is the {\em scepticism limit} \citep{matthews2018}.  Note that
\eqref{eq:S} holds for any value of $\alpha$, not just for the
traditional 5\% level.  The sufficiently sceptical prior variance
$\tau^2$ can be derived from \eqref{eq:S} and expressed as a function
of the variance $\sigma_o^2$ (the squared standard error, assumed to
be known) of the estimate $\hat \theta_o$, the corresponding test
statistic $t_o=\hat \theta_o/\sigma_o$ and $z_{\alpha/2}$, the
$1-\alpha/2$ quantile of the standard normal distribution \hl{\citep[Appendix]{held2019}}:
\begin{equation}\label{eq:tau2}
\tau^2 = \frac{\sigma_o^2}{t_o^2/z_{\alpha/2}^2 - 1},
\end{equation}
where $t_o^2>z_{\alpha/2}^2$ holds due to significance of the original study at level
$\alpha$. 

Equation \eqref{eq:tau2} 
shows that the sufficiently sceptical prior variance $\tau^2$ can be both smaller or larger
than $\sigma_o^2$, depending on the value of $t_o^2$. For a
``borderline'' significant result where $t_o^2$ is close to
$z_{\alpha/2}^2$, the sufficiently sceptical prior variance will be relatively large. 
If $t_o^2$ is
substantially larger than $z_{\alpha/2}^2$, then the sufficiently sceptical prior variance
will be relatively small.

\begin{center}
\begin{figure}[!h]
\begin{center}
\begin{knitrout}
\definecolor{shadecolor}{rgb}{0.969, 0.969, 0.969}\color{fgcolor}
\includegraphics[width=\maxwidth]{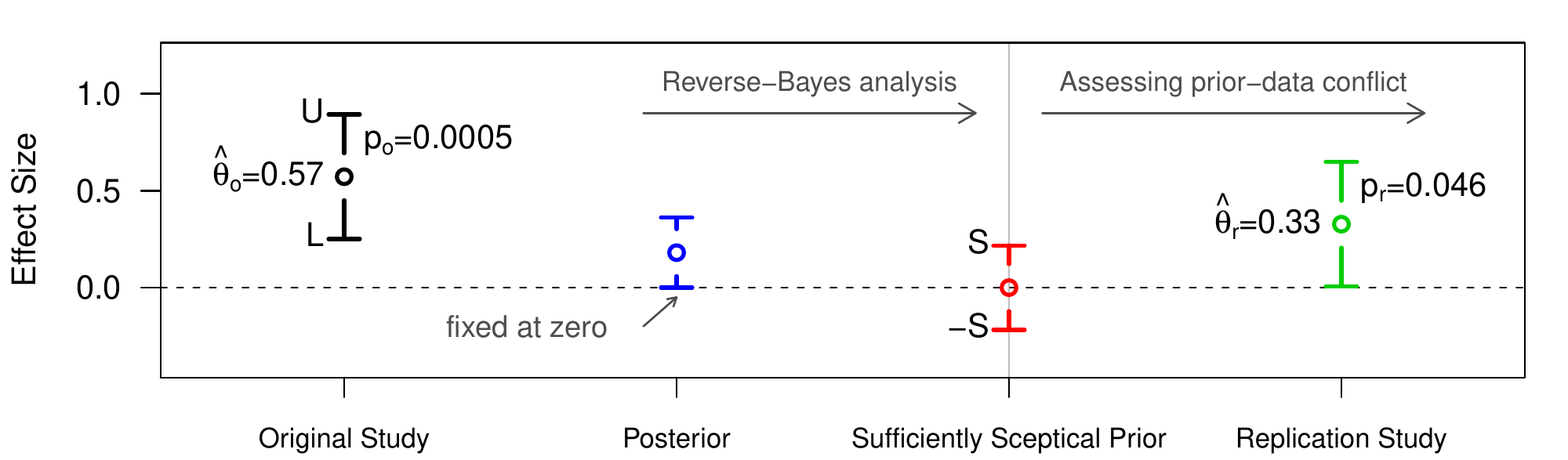} 

\end{knitrout}
\caption{Example of the assessment of replication success. The original study has effect estimate  $\hat \theta_o=
  0.57$ 
  (95\% CI from $L=0.25$ to $U=0.89$) and two-sided $p$-value 
  $p_o = 0.0005$.  
  The left part of the figure illustrates the \hl{reverse-Bayes derivation of the sufficiently sceptical prior with scepticism limit $S=0.22$
    based on}
  the original study result and the posterior with lower credible limit fixed at zero. \hl{The comparison} of the sufficiently sceptical prior with the replication study result ($\hat \theta_r = 0.33$, 95\% CI from 0.01 to 0.65, $p_r = 0.046$) 
  in the right part of the figure \hl{is used to assess potential prior-data conflict.}
  \label{fig:fig1}}

\end{center}
\end{figure}
\end{center}

\hl{The left part of} Figure \ref{fig:fig1} shows an example of this procedure. The original
study has effect estimate $\hat \theta_o=
0.57$ (95\% CI from
$L=0.25$ to $U=0.89)$ and
two-sided $p$-value $p_o = 0.0005$.  The scepticism
limit, calculated from \eqref{eq:S}, turns out to be
$S=0.22$.

. 

\subsection{\hl{Assessing Prior-Data Conflict}}
\label{sec:priorDataConflict}
The replication study \hl{shown in the right part of Figure \ref{fig:fig1} has} an effect estimate of
$\hat \theta_r = 0.33$ (95\% CI from
0.01 to 0.65, $p_r =
0.046$).  If the replication result is in conflict
with the sufficiently sceptical prior, the original result is deemed
credible. A visual comparison of the
sufficiently sceptical prior with the replication study result in the right part of Figure
\ref{fig:fig1} \hl{can be useful} to assess
potential conflict, but in general a more principled statistical
approach is needed.

One option is to 
consider the original study as credible, if the absolute value of the
effect estimate $\hat \theta_r$ from the replication study is larger
than the scepticism limit $S$ \citep{matthews:2001,matthews:2001b}. In the above example the effect estimate in
the replication study
($\hat \theta_r =
0.33$) is larger than the
scepticism limit ($S=0.22$), so the original study would be
considered credible at the 5\% level.
However, a disadvantage of this approach is that it does not take the
\hl{(known)} variance $\sigma_r^2$ of the replication estimate $\hat
\theta_r$ (in the following also assumed to be normally distributed)
into account. To address this issue, I propose to quantify prior-data
conflict based on the prior-predictive distribution of $\hat
\theta_r$, a normal distribution with mean zero and variance $\tau^2 +
\sigma_r^2$ \citep[Section
5.8]{sam:2004}. 
This \hl{is the established way to check the compatibility of prior and
  data \citep{box:1980,evans.moshonov2006} and}
  leads to the test statistic
\begin{equation}\label{eq:tbox}
t_{\mbox{\scriptsize Box}} = \frac{\hat \theta_r}{\sqrt{\tau^2+\sigma_r^2}}
\end{equation}
and the tail probability 
  $p_{\mbox{\scriptsize Box}} =\Pr(\chi^2(1) \geq t_{\mbox{\scriptsize Box}}^2)$
  as the corresponding upper tail of a
  $\chi^2$-distribution with one degree of freedom.  Small values of
  $p_{\mbox{\scriptsize
      Box}}$ indicate a conflict between the sufficiently sceptical prior and the
  estimate from the replication study and I define {\em replication
    success} at level $\alpha$ if $p_{\mbox{\scriptsize Box}} \leq
  \alpha$\hl{, or equivalently $t_{\mbox{\scriptsize Box}}^2 \geq z_{\alpha/2}^2$}.

In the example shown in Figure \ref{fig:fig1}, the
prior-predictive assessment of conflict gives
$t_{\mbox{\scriptsize Box}}= 1.65$ with Box's tail
probability $p_{\mbox{\scriptsize Box}}=0.098 > 0.05$, 
so the replication study is not successful
at the 5\% level, although both
the original and the replication study are significant at that
level. This illustrates that replication success is a more stringent
criterion than significance alone. For 
$\alpha=10\%$, Box's tail probability is somewhat
smaller ($p_{\mbox{\scriptsize Box}}=0.078$), and
we can declare replication success at the 10\% level.

The example illustrates how Box's tail probability 
can be used to assess replication success at
level $\alpha$. However, it is difficult to interpret the actual value of 
$p_{\mbox{\scriptsize Box}}$ as it depends on the choice of $\alpha$.
Furthermore, assessment of replication success is only possible if the
original study result is significant at level $\alpha$ as otherwise
the sufficiently sceptical prior would not exist and $p_{\mbox{\scriptsize Box}}$ could not be computed. 
These issues
motivate the work described in the next section where I introduce the
sceptical $p$-value, a \hl{quantitative} measure for replication success that is
independent of the level $\alpha$.

\section{The Sceptical $p$-Value}
\label{sec:sceptical.p}

Instead of dichotomizing replication
studies into successful yes/no at some arbitrary \hl{level} $\alpha$,
I now propose the {\em sceptical $p$-value} $p_S$ to assess
replication success quantitatively.  The idea is to determine the
largest confidence level $1-p_S$ for the \hl{original} confidence interval, \hl{at which} we are able to declare replication
success at level $p_S$. This parallels the
duality of ordinary $p$-values and confidence intervals, where the
largest confidence level $1-p$ \hl{at which} we are able to declare
significance can be used to compute the ordinary $p$-value $p$.
Replication success at any pre-specified level $\alpha$ is then 
equivalent to $p_S \leq \alpha$, \hl{just as significance at level
$\alpha$ is equivalent to $p \leq \alpha$.}

To determine $p_S$, let $c = \sigma_o^2/\sigma_r^2$ denote the
ratio of the variances of the original and replication
effect estimates and let $t_r=\hat \theta_r/\sigma_r$ denote the test statistic of the replication
study. With
\eqref{eq:tau2} we can derive the prior-predictive variance of $\hat \theta_r$:
\begin{equation}\label{eq:priorPredVar}
\tau^2 + \sigma_r^2 = \sigma_r^2 \left(\frac{c}{t_o^2/z_{\alpha/2}^2 - 1} + 1 \right).
\end{equation}
Using \eqref{eq:tbox} and \eqref{eq:priorPredVar}, the requirement
$t_{\mbox{\scriptsize Box}}^2 = \hat \theta_r^2/(\tau^2 + \sigma_r^2) \geq z_{\alpha/2}^2$ for 
replication success \hl{at level $\alpha$} 
\hl{can be written as}
\begin{equation}\label{eq:extrinsic.p}
\left({t_o^2}/{z_{\alpha/2}^2}-1 \right) \left({t_r^2}/{z_{\alpha/2}^2} - 1 \right) \geq c, 
\end{equation}
see Appendix \ref{app:app1} for a derivation. 
Significance of
the original study implies that $z_{\alpha/2}^2 < t_o^2$ holds,
therefore $z_{\alpha/2}^2 < t_r^2$ must also hold to ensure that the left hand
side of equation \eqref{eq:extrinsic.p} is positive. 
The required squared quantile $z_S^2=z^2_{p_S/2}$ to obtain equality in
\eqref{eq:extrinsic.p} \hl{must therefore fulfill 
\begin{equation}\label{eq:z2.range}
  0 \leq z_S^2 < \min\{t_o^2, t_r^2\}
\end{equation}}
and defines the sceptical $p$-value
 $p_S = 2 \left[1-\Phi({z_S})\right]$ via
\begin{equation}\label{eq:extrinsic.p2}
\left({t_o^2}/{z_S^2}-1 \right) \left( {t_r^2}/{z_S^2} - 1 \right) = c .
\end{equation}
The requirement $p_S \leq \alpha$ for replication success at level $\alpha$ \hl{now} translates to
$z_S^2 \geq z_{\alpha/2}^2$. 

Equation \eqref{eq:extrinsic.p2} can be re-written as
\begin{equation}\label{eq:quadEq}
(c-1) z_S^4 + 2 z_S^2 \, t^2_{A} = t^2_A \, t^2_H ,
\end{equation}
where $t^2_A=(t_o^2 + t_r^2)/2$ is the arithmetic and $t^2_H=2/(1/t_o^2+
1/t_r^2)$ the harmonic mean of the squared test statistics $t_o^2$ and $t_r^2$. 
The only solution of  \eqref{eq:quadEq} that fulfills \hl{\eqref{eq:z2.range}} 
is  \begin{eqnarray}
z_S^2 
&=&  
\left\{
\begin{array}{ll}
{t_H^2}/{2} & \mbox{ for } c=1 \mbox{ and } \\
\frac{1}{c-1} 
\left\{
\sqrt{t_A^2 \left[{t_A^2} + (c-1) t^2_H\right]} - {t_A^2} 
\right\} & \mbox{ for } c \neq 1. \label{eq:solution}
\end{array}
\right.
\end{eqnarray}

In the introductory example the original and the replication
confidence intervals have the same width, so $c=1$ and $z_S^2$ is
simply half the harmonic mean of $t_o^2=12.19$ and
$t_r^2=3.99$, \ie
$z_S^2 = 3.00$,
$\abs{z_S} = 1.73$ and
$p_S=2 \left[1-\Phi(1.73)\right] =
0.083$.  We can thus declare replication
success at any \hl{pre-specified} level $\alpha \geq 0.083$.

\subsection{Properties}
\label{sec:properties}
Inequation 
\hl{\eqref{eq:z2.range}} implies that the
sceptical $p$-value $p_S$ is always larger than both the
original and the replication $p$-values $p_o$ and $p_r$. 
Closer inspection of equation \eqref{eq:extrinsic.p2} shows 
that $z_S^2$ is increasing with increasing $t_o^2$ (for
fixed $t_r^2$ and $c$) and also with increasing $t_r^2$ (for fixed
$t_o^2$ and $c$). Therefore, the smaller $p_o$ (or $p_r$), the smaller
$p_S$ (for fixed $c$). Furthermore, for fixed test statistics $t_o$ and
$t_r$ (so fixed $p$-values $p_o$ and $p_r$), the solution $z_S^2$ of
\eqref{eq:extrinsic.p2} will decrease with increasing variance ratio
\[
  c = \frac{\sigma_o^2}{\sigma_r^2} = \frac{t_r^2/t_o^2}
  {\hat\theta_r^2/\hat\theta_o^2}
\]
Since $t_r^2/t_o^2$ is fixed, $c$ increases with decreasing 
squared effect size ratio ${\hat \theta_r^2}/{\hat \theta_o^2}$.
In other words, for the same ordinary 
$p$-values $p_o$ and $p_r$, the sceptical $p$-value 
$p_S$ increases with \hl{decreasing} absolute replication effect
estimate relative to the original effect estimate. This is a
desired property, as replication studies with smaller effect
estimates than the original estimates are considered less credible 
\citep{Simonsohn2015}.

To illustrate the dependence of $p_S$ on the variance ratio
$c$, consider a scenario where $p_o=p_r=0.01$, so
$t_o^2=t_r^2$ and therefore $c = {\hat \theta_o^2}/{\hat
  \theta^2_r}$.  First assume equal effect sizes ${\hat
  \theta_r}={\hat \theta_o}$, so $c=1$.  The sceptical
$p$-value turns out to be $p_S=0.069$. For ${\hat
  \theta_r}= {\hat \theta_o}/2$
($c=4$) we obtain a larger value
($p_S=0.14$) because the effect
estimate $\hat
\theta_r$ of the replication study is just half as large as the
original estimate $\hat \theta_o$.  On the other hand, for ${\hat
  \theta_r}=2 \, {\hat \theta_o}$ ($c=1/4$) the sceptical
$p$-value gets smaller
($p_S=0.035$).  This asymmetry in the incorporation
of the original and replication study data is natural, placing less
weight on replication studies with relatively small effect
estimates. This is the case in the introductory example, where 
substantial shrinkage of the replication effect estimate 
leads to a relatively 
large sceptical $p$-value.

It is also interesting to study limiting values of the sceptical
$p$-value.  
If we let $\sigma_o^2 \downarrow 0$ for fixed $\hat \theta_o \neq 0$, 
\eqref{eq:quadEq} reduces to the requirement $z_S^2 = t_r^2$, as shown in Appendix \ref{app:app2}. 
Thus, the ordinary $p$-value of the replication study \hl{is} a
special case of the sceptical $p$-value if the uncertainty
of  the original effect estimate is ignored. 
On the other hand, ignoring the uncertainty of $\hat \theta_r \neq 0$ via
$\sigma_r^2 \downarrow 0$ leads to $z_S^2 \downarrow z_M^2$ where
\begin{equation}\label{eq:p.matthews}
  z_M^2 = \frac{\sqrt{d \, (d+4)} - d}{2} \, t_o^2,
\end{equation}
with $d= \hat \theta_r^2/\hat \theta_o^2$, see Appendix \ref{app:app2}
for a proof.  Using the criterion $z_M^2 \geq z_{\alpha/2}^2$ rather
than $z_S^2 \geq z_{\alpha/2}^2$ to assess replication success
corresponds to the \citet{matthews:2001,matthews:2001b} approach \hl{mentioned in Section \ref{sec:priorDataConflict}}. For any value of $d$, 
$z_M^2$ is smaller than $t_o^2$ but can be larger than $t_r^2$. Ignoring the uncertainty of the replication effect estimate 
may thus lead to the declaration of replication success, even if the replication
study is not conventionally significant on its own.

We may also consider the case $c \downarrow 0$ for fixed $t^2_o$ and
$t^2_r$, where \eqref{eq:solution} increases with limit
\begin{equation}
z_S^2 \uparrow \min\{t^2_o, t^2_r \}, 
  \label{eq:eq1} 
\end{equation}
as shown in Appendix \ref{app:app3}. Therefore $p_S \downarrow \max\{p_o, p_r\}$ 
for $c \downarrow 0$, which we will use in Section \ref{sec:sec4}. 

\subsection{Relationship to Intrinsic Credibility}
\label{sec:intrinsic}
The concept of intrinsic credibility has been proposed in
\cite{matthews2018} to check the credibility of "out of the blue"
findings without any prior support.  In the present context this
corresponds to an original study in the absence of a replication
study. The idea is to evaluate the credibility of the original
study if we would be able to observe exactly the same result in the
replication study.

The approach by \cite{matthews2018} corresponds to the case where we ignore the uncertainty of the 
(hypothetical) replication study and thus leads to 
\eqref{eq:p.matthews} with $d=1$:
  $z_M^2 = ({\sqrt{5} - 1})/{2} \, t_o^2 \approx 0.618 \, t_o^2$.
However, if we incorporate the uncertainty using the
prior-predictive approach by \cite{box:1980}, then we obtain $z_S^2 = 0.5 \, {t_o^2}$
as a special case of \eqref{eq:solution}
for $c=1$ and $t_o^2=t_r^2$. 
Now $p_S$ reduces to the $p$-value for intrinsic credibility,
\begin{equation}\label{eq:eq3}
  p_{IC} = 2\left[1-\Phi\left(t_o/\sqrt{2} \right)\right], 
\end{equation}
as
proposed in \cite{held2019} for the assessment of claims of new discoveries. Intrinsic credibility at level
$\alpha$ is achieved if $p_{IC} \leq \alpha$, \ie 
$t_o^2 \geq 2 \, z_{\alpha/2}^2$, 
which is equivalent to
$p_o \leq \alpha_{IC}$, where
\begin{eqnarray}\label{eq:eq2}
\alpha_{IC} &=& 2 \left\{1 - \Phi\left(\sqrt{2} \, z_{\alpha/2} \right) \right\} \label{eq:iP}
\end{eqnarray}
is the $p$-value threshold for intrinsic credibility. For $\alpha=0.05$ we obtain 
$\alpha_{IC} = 0.0056$, for $\alpha=0.10$ we have
$\alpha_{IC} = 0.02$. These thresholds will become important in Section 
\ref{sec:design}. 

\subsection{One-Sided Sceptical $p$-Values}
\label{sec:sec2.4}
The procedure described above is designed for standard two-sided confidence
intervals and assesses replication success in a two-sided fashion, as the sign
of $\hat \theta_r$ does not matter in the computation of the sceptical
$p$-value. In extreme cases, it may therefore happen that a replication study is
classified as successful although the signs of $\hat \theta_o$ and
$\hat \theta_r$ differ. 
This ``replication paradox'' may also occur in a Bayes factor  
approach, see \citet{Ly_etal2018} for details. 

It is therefore of interest to adapt the sceptical $p$-value to the
one-sided setting. Without loss of generality consider the one-sided
alternative $H_1$: $\theta > 0$ to $H_0$: $\theta=0$ and assume, that
$\hat \theta_o > 0$.  \hl{We now start with
a one-sided confidence interval
for $\theta$ at level $1-\tilde \alpha$ whose lower limit $\hat \theta_o - z_{\tilde \alpha} \, \sigma_o$ 
equals the lower limit $L$ of the corresponding two-sided confidence interval at
level $1-2 \tilde \alpha$.} The variance \hl{$\tau^2$} of the
\hl{sufficiently} sceptical prior therefore is \eqref{eq:tau2}
with $z_{\alpha/2}$ replaced by $z_{\tilde \alpha}$.

The \hl{obvious} one-sided requirement for replication success
$t_{\mbox{\scriptsize Box}}$
$\hl{=\hat \theta_r / \sqrt{\tau^2+\sigma_r^2}} \geq z_{\tilde
  \alpha}$ now \hl{replaces the two-sided requirement
  $t_{\mbox{\scriptsize Box}}^2 \geq z_{\alpha/2}^2$ and} ensures that
the replication paradox cannot occur.  \hl{If $\hat \theta_r \geq 0$,
  we can hence still use \eqref{eq:solution} to compute $z_S^2$ from
  $t_o$, $t_r$ and $c$ and the one-sided sceptical $p$-value turns out
  to be $\tilde p_S = 1-\Phi({z_S})$, so half of the two-sided
  sceptical $p$-value: $\tilde p_S = p_S/2$.  If the replication
  effect estimate is in the wrong direction, \ie $\hat \theta_r < 0$,
  it is natural to set $\tilde p_S = 1 - p_S/2$.} \hl{The same
  relationship holds between ordinary one- and two-sided $p$-values,
  of course, which implies that the one-sided sceptical $p$-value
  $\tilde p_S$ is always larger than the ordinary one-sided $p$-values
  $\tilde p_o$ and $\tilde p_r$ from the two studies. }

One-sided sceptical $p$-values are appropriate if the study protocol
of the original study is already formulated in a one-sided fashion. 
A post-hoc (after the original
study result is known) formulation of a one-sided alternative would
require halving the original two-sided significance level $\alpha$ to
$\tilde \alpha = \alpha/2$. The one-sided assessment of replication success 
\hl{at level $\tilde \alpha$}
is then equivalent to the two-sided procedure at \hl{the original} level $\alpha$, if the
signs of the original and replication effect estimates agree. This
procedure ensures that the replication paradox cannot occur. 

\hl{
However, the one-sided $p$-value mapping (from
$\tilde p_o$ and $\tilde p_r$ to $\tilde p_S$)
will be different from the corresponding two-sided mapping (from
$p_o$ and $p_r$ to $p_S$) 
because the same ordinary one- and two-sided 
$p$-values correspond to different test statistics $t_o$ and $t_r$. 
For numerical illustration suppose that
$\tilde p_o = \tilde p_r = 0.01$ (so $p_o=p_r=0.02$ and
$t_o=t_r=2.33$) and that $\hat \theta_o$ and
$\hat \theta_r$ are both positive. Then
$\tilde p_S = 0.05$ for $c=1$,
$\tilde p_S = 0.09$ for $c=4$ and
$\tilde p_S = 0.029$ for $c=1/4$. These one-sided
sceptical $p$-values are slightly smaller than the two-sided sceptical
$p$-values for two-sided $p_o=p_r=0.01$ (where $t_o=t_r=2.58$), as reported in Section
\ref{sec:properties} ($p_S=0.069$, $0.14$ and $0.035$ for 
$c=1$, $4$ and $1/4$, respectively).  This illustrates that the one-sided assessment of
replication success based on one-sided ordinary $p$-values is 
slightly less stringent than the two-sided assessment based on two-sided
ordinary $p$-values, if the original and replication effect estimates have the same sign.}

\subsection{The Distribution Under The Null}
\label{sec:sec4}

It is interesting to compare the distributions of $p_o$ (or $p_r$), 
$p_{IC}$ and $p_S$ 
under the assumption of no effect, \hl{where} 
the ordinary $p$-value is uniformly distributed. We can easily
derive the density of $p_{IC}$ with a
change-of-variables using \eqref{eq:eq3}:
$f(p_{IC}) = 2 \sqrt{\pi} \, \varphi\left\{t(p_{IC})\right\}$,
here $\varphi(.)$ is the standard normal density function and
$t(p_{IC})=\Phi^{-1}(1- p_{IC}/2)$.

\begin{figure}[!ht]
\begin{center}
\begin{knitrout}
\definecolor{shadecolor}{rgb}{0.969, 0.969, 0.969}\color{fgcolor}
\includegraphics[width=\maxwidth]{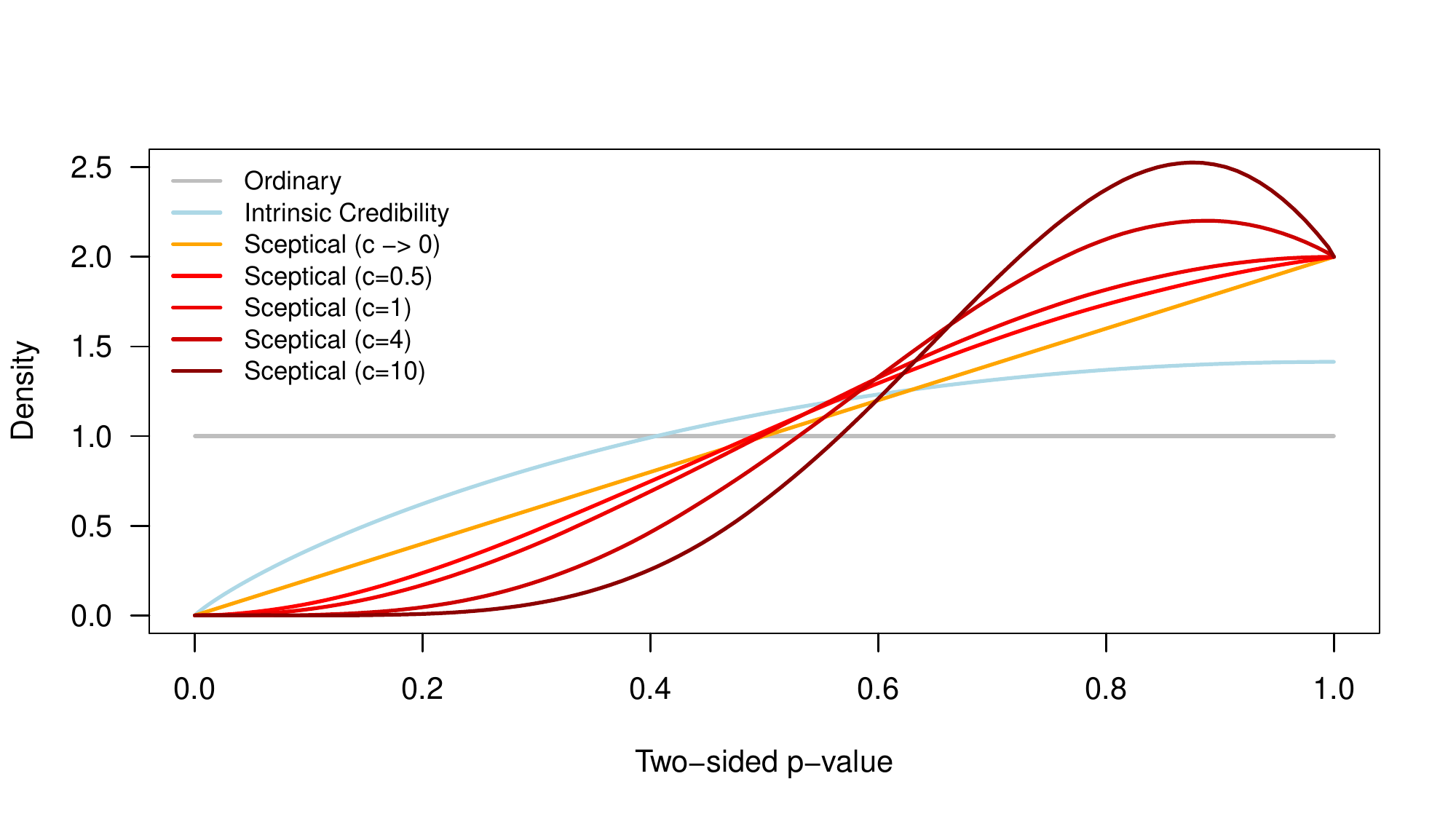} 

\end{knitrout}
\end{center}
\caption{The density function of the sceptical $p$-value for
  different values of the variance ratio $c$ under the assumption of
  no effect.  The density for the
  limiting case $c \rightarrow 0$ as well as the
  density of the $p$-value for intrinsic credibility and the ordinary $p$-value are also shown.}
\label{fig:fig5}
\end{figure}

The distribution of $p_S$ can be studied via stochastic simulation.  Density
estimates are displayed in Figure \ref{fig:fig5} for different values
of the variance ratio $c$ based on \ensuremath{5\times 10^{6}} samples each.  We can see that 
the risk of small ``false positive''
sceptical $p$-values is drastically reduced, compared to ordinary
$p$-values based on one study only. Note
that the variance is usually inversely proportional to the sample size
of each study, \ie $\sigma_o^2 = \kappa^2/n_o$ and
$\sigma_r^2 = \kappa^2/n_r$ for some unit variance $\kappa^2$, say.
Then $c = n_r/n_o$, so the variance factor $c$ is increasing with
increasing sample size $n_r$ of the replication study.  The
distribution of $p_S$ in Figure \ref{fig:fig5} is shifted to the right
with increasing $c$, so an increasing sample size of the replication
study reduces the risk of a false claim of replication success.

From \eqref{eq:eq1} we know that for $c \downarrow 0$ we have $p_S \downarrow \max\{p_o, p_r\}$,
which follows a triangular $\Be(2, 1)$ distribution if $p_r$ and $p_o$ are independently uniform.  The corresponding density function 
is shown in  Figure \ref{fig:fig5}, as well as
the density function of $p_o$ and $p_{IC}$. The triangular distribution 
gives the upper bound $\alpha^2$ for the tail probability
$\Pr(p_S \leq \alpha \given H_0)$ for sufficiently small
$\alpha$ and any value of the variance ratio $c$. For example,
for $\alpha=0.05$ we obtain $\Pr(p_S \leq 0.05 \given H_0) \leq
0.0025$ for any $c$. This is to be compared with $\Pr(p_o \leq 0.05 \given H_0) = 0.05$
and 
$\Pr(p_{IC} \leq 0.05 \given H_0) = 0.0056$. 
However, $\alpha^2$ is not a particularly sharp bound. If, for example,
the replication sample size equals the original sample size 
($c=1$), then $\Pr(p_{S} \leq 0.05 \given H_0) \approx 0.0001$
is much smaller than 0.0025.

\section{Power and Sample Size Calculations}
\label{sec:design}

Replication success is not only a function of the two $p$-values from
the original and replication study, but also of sample size, which
enters in the variance ratio $c$.  The computation of the power or the required
\hl{replication} sample size \hl{to achieve replication success} is hence
more challenging than in standard sample size calculations.  
A larger sample size will be required since replication success (defined as $p_S \leq \alpha$) implies
significance of the replication study ($p_r \leq \alpha$). Furthermore, 
the required sample size will depend on the $p$-value
$p_o$ from the original study. 

\hl{The Bayesian assessment of sample size uses a design prior
  \citep{OHaganStevens2001,OHagan2005} to express prior beliefs about
  the true effect size.  In order to power a study for replication success,} the
results from the original study will thus enter \hl{in two ways}, as
\hl{design} prior for the effect size and in the subsequent assessment
of replication success. For the former I will distinguish two cases, a
normal prior with mean
$\hat \theta_o$ and variance $\sigma_o^2$ and a point prior at $\hat \theta_o$. The normal prior incorporates the
uncertainty of $\hat \theta_o$ while the point prior does not, \hl{in
analogy to the concepts of predictive and conditional power in
clinical trials \citep{SpiegelhalterFreedman1986,spiegelhalter1986}}.

Suppose $n_o$ is the size of the original study sample and $n_r$ the
sample size of the replication study, so $\sigma_o^2=\kappa^2/n_o$ and
$\sigma_r^2=\kappa^2/n_r$, where $\kappa^2$ is the unit variance from
one observation. Then $c=n_r/n_o$, which would also hold in a balanced
two-sample design with respective sample sizes $n_o$ and $n_r$ per
group. 
Under an initial uniform prior for $\theta$, the sampling distribution $\hat
\theta_o \sim \Nor(\theta, \sigma_o^2)$ of the original study now serves
as prior distribution 
$
\theta \given \hat
\theta_o \sim \Nor\left(\hat \theta_o, \sigma_o^2 \right)
$
with prior-predictive distribution 
\begin{equation}\label{eq:priorPredictive}
\hat \theta_r \given \hat \theta_o \sim \Nor\left(\hat
\theta_o, \sigma_o^2 + \sigma_r^2 = \kappa^2 \left\{\frac{1}{n_o} + \frac{1}{n_r} \right\}\right)
\end{equation}
for the observed effect $\hat \theta_r$ in the replication study.
Then $t_r^2$ follows a scaled non-central $\chi^2$-distribution with one
degree of freedom, scaling factor \hl{$1+c$} 
and non-centrality parameter
\hl{$t_o^2/(1+1/c)$}, 
as shown in Appendix \ref{app:app4}. For the alternative point prior at $\theta=\hat \theta_o$, 
$t_r^2$ follows a non-central $\chi^2$-distribution with one degree of freedom and non-centrality parameter 
\hl{$\lambda = c \, t_o^2$}.

To compute the power for replication success, the \hl{relative sample size} $c = n_r/n_o$ in \hl{\eqref{eq:solution}}
is fixed.  Then $z_S^2$ and the sceptical
$p$-value $p_S$ are monotone functions of $t_r^2$ and we can compute
the power for replication success at any level $\alpha$.  We can also
calculate the required \hl{relative sample size $c=n_r / n_o$}
at some pre-defined power for replication success. Both \hl{tasks} require
application of numerical root-finding algorithms. Computational details are omitted
here.

\subsection{Power calculations}

\begin{figure}[!ht]
\begin{center}
\begin{knitrout}
\definecolor{shadecolor}{rgb}{0.969, 0.969, 0.969}\color{fgcolor}
\includegraphics[width=\maxwidth]{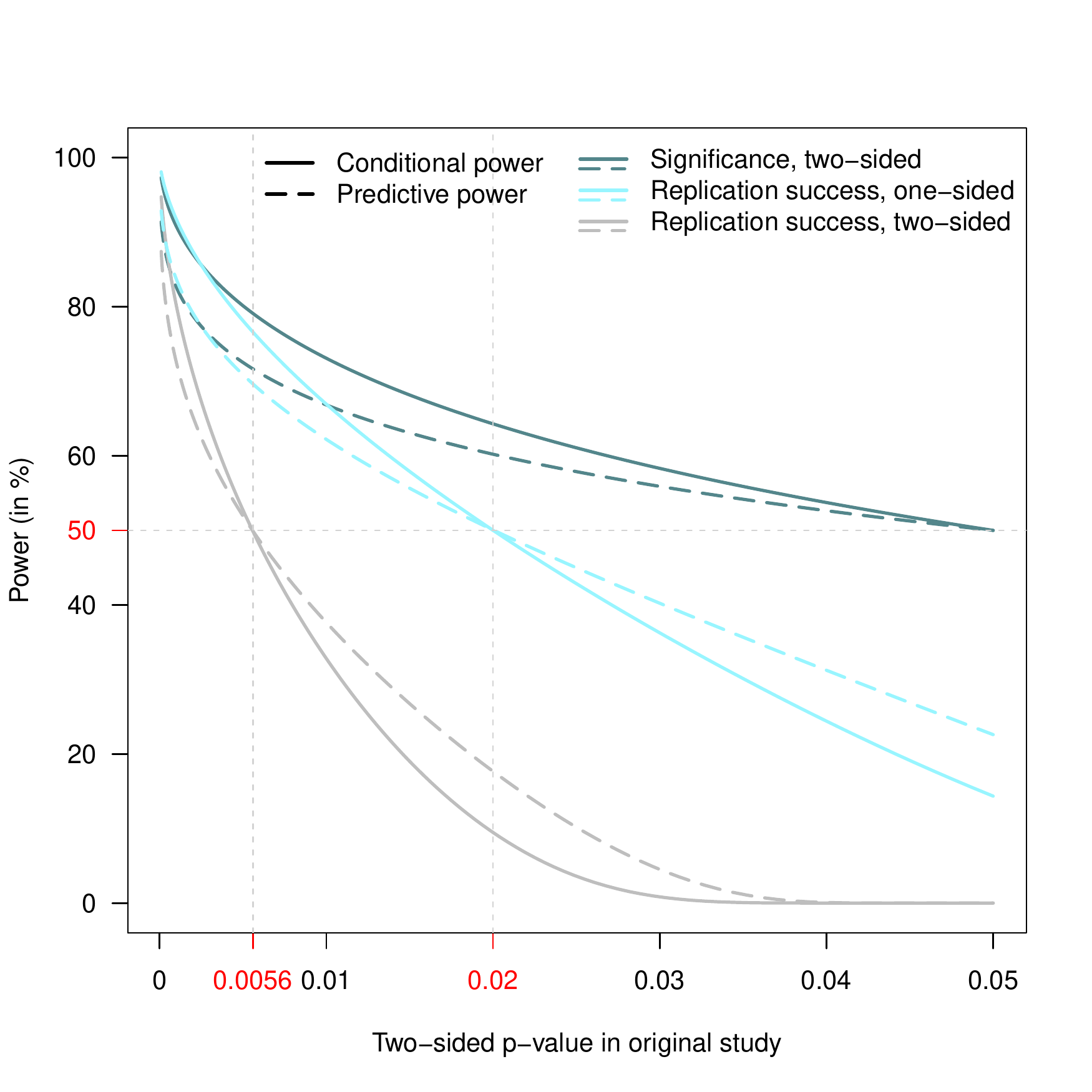} 

\end{knitrout}
\end{center}
\caption{Power calculations for a replication study with sample size
  equal to the original study \hl{($c=1$)}. Shown is \hl{conditional and predictive} power for
  significance \hl{(two-sided)} and for replication success \hl{(one-
    and two-sided)} at level $\alpha=5\%$ as a function of the \hl{two-sided}
  $p$-value of the original study. The one- and two-sided thresholds for
  intrinsic credibility are marked in red on the $x$-axis.}
\label{fig:fig6}
\end{figure}

Figure \ref{fig:fig6} compares the power for significance with the
power for replication success for a replication study with sample size
equal to the original study \hl{($c=1$)} at level $\alpha=5\%$ as a
function of the \hl{two-sided} $p$-value $p_o$ of the original study.  Power calculations for significance 
aim
to detect the effect estimate $\hat \theta_o$ from the original
study with a standard two-sided significance test. Not accounting for the
associated uncertainty corresponds to 
\hl{the concept of conditional power}, whereas \hl{predictive power
  calculations are based on} a normal \hl{design prior} with mean
$\hat \theta_o$ and variance $\sigma_o^2$ \citep[Equation
6.4]{sam:2004}.  The results are in accordance with those reported in
\cite{Goodman1992}. In particular, for $p_o=0.05$ the power is 50\%
\hl{both for conditional and predictive power with conditional power increasing
faster than predictive power
for smaller $p$-values $p_o$.}

\hl{Conditional and predictive} power for replication success is
also shown in Figure \ref{fig:fig6} for two-sided $\alpha = 5\%$ \hl{respectively one-sided $\tilde \alpha = 5\%$}.  
As expected, the power for
replication success is lower than for significance, and drops quickly
to zero for values of $p_o$ close to 0.05. Remarkably, in the
two-sided case, a \hl{conditional and predictive} power of 50\% is
attained at $p_o=0.0056$. This is the threshold \eqref{eq:eq2} for
intrinsic credibility at level $\alpha=5\%$, as described in Section
\ref{sec:intrinsic}.  In the one-sided case a power of 50\% is
obtained at $p_o=0.02$, the threshold for intrinsic credibility at
two-sided level $\alpha=10\%$. Therefore, only intrinsically credible
results (based on the threshold \eqref{eq:eq2})
ensure that the power for success of an identically designed
replication study exceeds 50\%. This intriguing feature highlights the
difficulty to achieve replication success if the evidence from the
original study is only suggestive and provides a new argument for \hl{more stringent} $p$-value thresholds for claims of new
discoveries \citep{Johnson2013,BenjaminEtAlinpress, Ioannidis2018, held2019}.

This surprising result can be explained as follows: If the non-centrality parameter
$\lambda$ of a non-central $\chi^2(1)$-distribution is reasonably large,
say $\lambda > 4$, then the median is approximately equal to
$\lambda$. Under the point prior and for $c=1$, the non-centrality
parameter of the distribution of $t_r^2$ is $\lambda=t_o^2$, so 
$\Med(t_r^2) \approx t_o^2$. The
sceptical $p$-value $p_S = 2 \left[1-\Phi({z_S})\right]$ is then defined  through
$z_S^2 = t_H^2/2 = (1/t_o^2 + 1/t_r^2)^{-1}$, 
so the median of $z_{S}^2$ is approximately $t_o^2/2$. 
Replication success is thus
achieved with 50\% probability for
$z_{\alpha/2}^2 = z_{S}^2 \approx t_o^2/2$, \ie
$t_o^2 \approx 2 \, z_{\alpha/2}^2$. This corresponds to the 
intrinsic credibility threshold \eqref{eq:eq2} for the ordinary 
$p$-value $p_o$ from the original study.

We obtain essentially the same result under the normal prior, where now 
$\lambda = t_o^2/2$, which combined with a scaling factor of 2 also
leads to $\Med(t_r^2) \approx t_o^2$ for sufficiently large $t_o^2$.
The rest of the argument is as
above, but note that this approximation is slightly less precise,
because the non-centrality parameter is half as large than under the point prior. 
The approximation is, however, still very good: For $\alpha=5\%$, the exact power for replication success at 
$p_o = 2[1-\Phi(\sqrt{2} \, z_{0.025})]$ is 50.00001\% for the point prior
and 50.00459\% for the normal prior.

\begin{figure}[!ht]
\begin{center}

\begin{knitrout}
\definecolor{shadecolor}{rgb}{0.969, 0.969, 0.969}\color{fgcolor}
\includegraphics[width=\maxwidth]{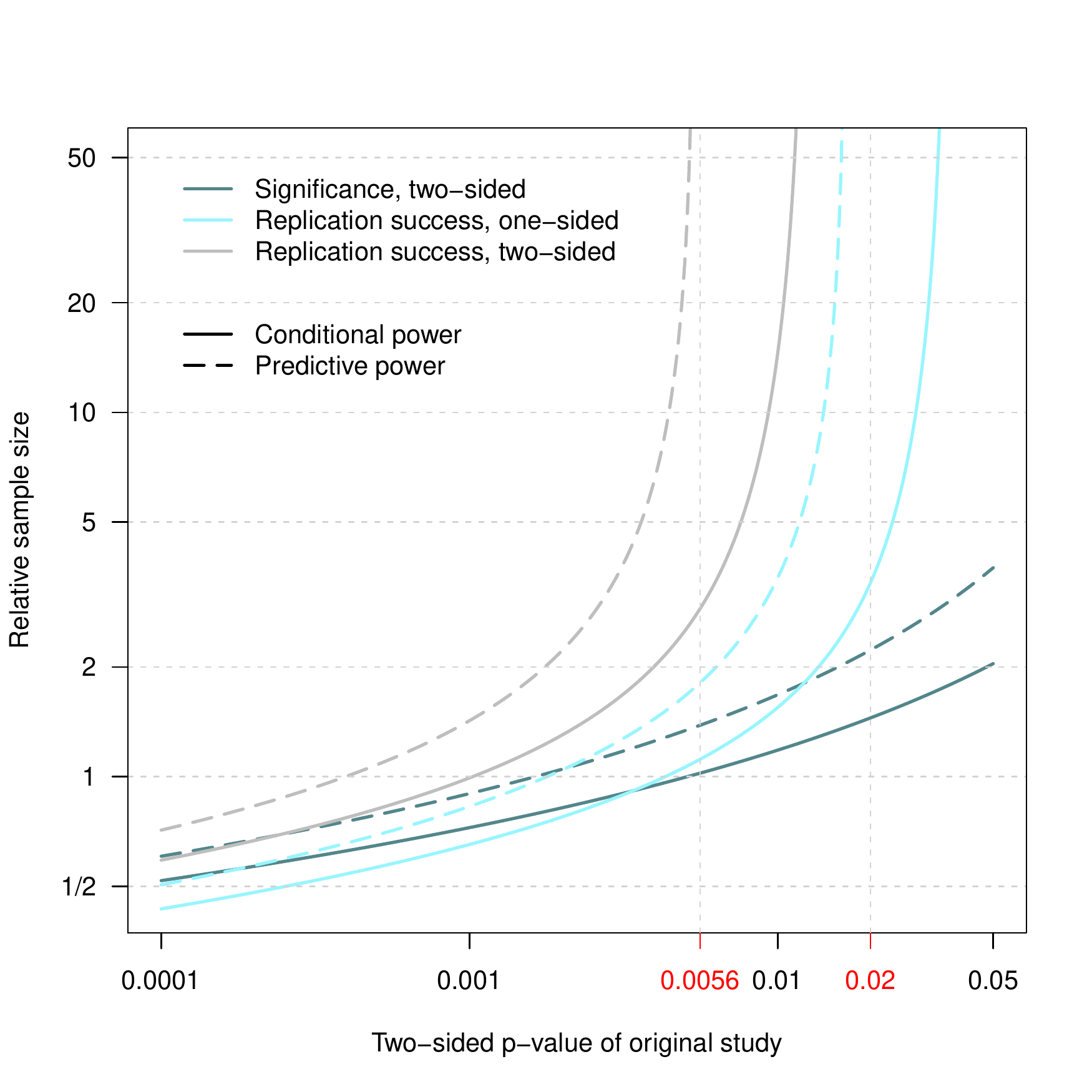} 

\end{knitrout}

\end{center}
\caption{\hl{Relative} sample size $c=n_r/n_o$ to achieve significance \hl{(two-sided)} and 
  replication success \hl{(one-
    and two-sided)} 
  with 80\% \hl{conditional or predictive} power for 
  $\alpha=0.05$
  as a function of the original two-sided $p$-value $p_o$. The one- and two-sided thresholds for
  intrinsic credibility are marked in red on the $x$-axis.}
\label{fig:fig7}
\end{figure}
\subsection{Sample size calculations}
Figure \ref{fig:fig7} compares different
\hl{strategies} to determine the \hl{replication} sample size for \hl{two-sided}
$\alpha=5$\% \hl{respectively one-sided $\tilde \alpha=5$\%} and original \hl{two-sided} $p$-values $p_o$ between
0.0001 and 0.05.
\hl{The power to achieve significance respectively replication success 
is fixed at 80\%.}

Standard (two-sided) sample size calculations \hl{based on conditional
  power} give \hl{relative} sample sizes between 0.52
and 2, depending on the $p$-value $p_o$ of
the original study.  Incorporating the uncertainty from the original
study \hl{based on predictive power} gives
\hl{relative} sample sizes between 0.61 and
3.7.

The required \hl{relative} sample size for two-sided replication success is
larger than the one for significance alone and depends \hl{more drastically on the $p$-value $p_o$ of
  the original study. \hl{First consider the case of conditional
    power}.  If $p_o$ is smaller than 0.001, the required relative
  sample size $c$ is smaller one, so the replication sample size $n_r$
  does not need to be larger than the original sample size $n_o$.}
  However, the required sample size explodes for larger $p$-values
  with an asymptote around
  $p_o=0.012$. This highlights the 
  difficulty to achieve 80\% power for replication success
  with original $p$-values between 0.01 and 0.05. Even larger
sample sizes are required \hl{based on predictive rather than
  conditional power} with an asymptote around 
  $p_o=0.005$. 
    
  The curves shift a little to the right when we assess replication
  success in a one-sided fashion, pushing the asymptotes towards
  $p_o=0.035$ 
  for conditional and
  $p_o=0.017$ 
  for predictive power.
  \hl{The predictive power asymptotes are remarkably close to the
    corresponding thresholds 0.0056 and 0.02 for
    intrinsic credibility, also shown in Figure \ref{fig:fig7}. Of
    course, the asymptotes would change for power values different
    from 80\%.}

\section{Replication Success in Psychological Science}
\label{sec:sec5}
I now re-analyse data from the \citet{OSC2015} project, \hl{a
  multi-year endeavour that replicated 100 scientific studies selected
  from three prominent psychology journals.} Effect sizes have been transformed to correlation coefficients
$\hat \rho$ where application of 
Fisher's $z$-transformation $\hat \theta = \tanh^{-1}(\hat \rho)$
justifies a normal
assumption with the standard error being a function only of the
\hl{nominal} study sample size $n$: $\se(\hat \theta) = 1/\sqrt{n-3}$,
\hl{so the effective sample size is $n-3$ with variance ratio
  $c=(n_r-3)/(n_o-3)$.}  Effective sample sizes are available for 73 studies, the so-called Meta-Analytic subset
\citep{Johnson2016}.  \hl{Two-sided $p$-values $p_o$ and $p_r$ have
  been calculated assuming normality of the corresponding test
  statistics $t_o = \hat\theta_o /\se(\hat \theta_o)$ and
  $t_r = \hat\theta_r /\se(\hat \theta_r)$, respectively.} \hl{I have
  not included the remaining 27 studies where a normal approxi\-mation
  is questionable since the standard errors of $\hat\theta_o$ and $\hat\theta_r$ are not available. }

\begin{figure}
\begin{knitrout}
\definecolor{shadecolor}{rgb}{0.969, 0.969, 0.969}\color{fgcolor}
\includegraphics[width=\maxwidth]{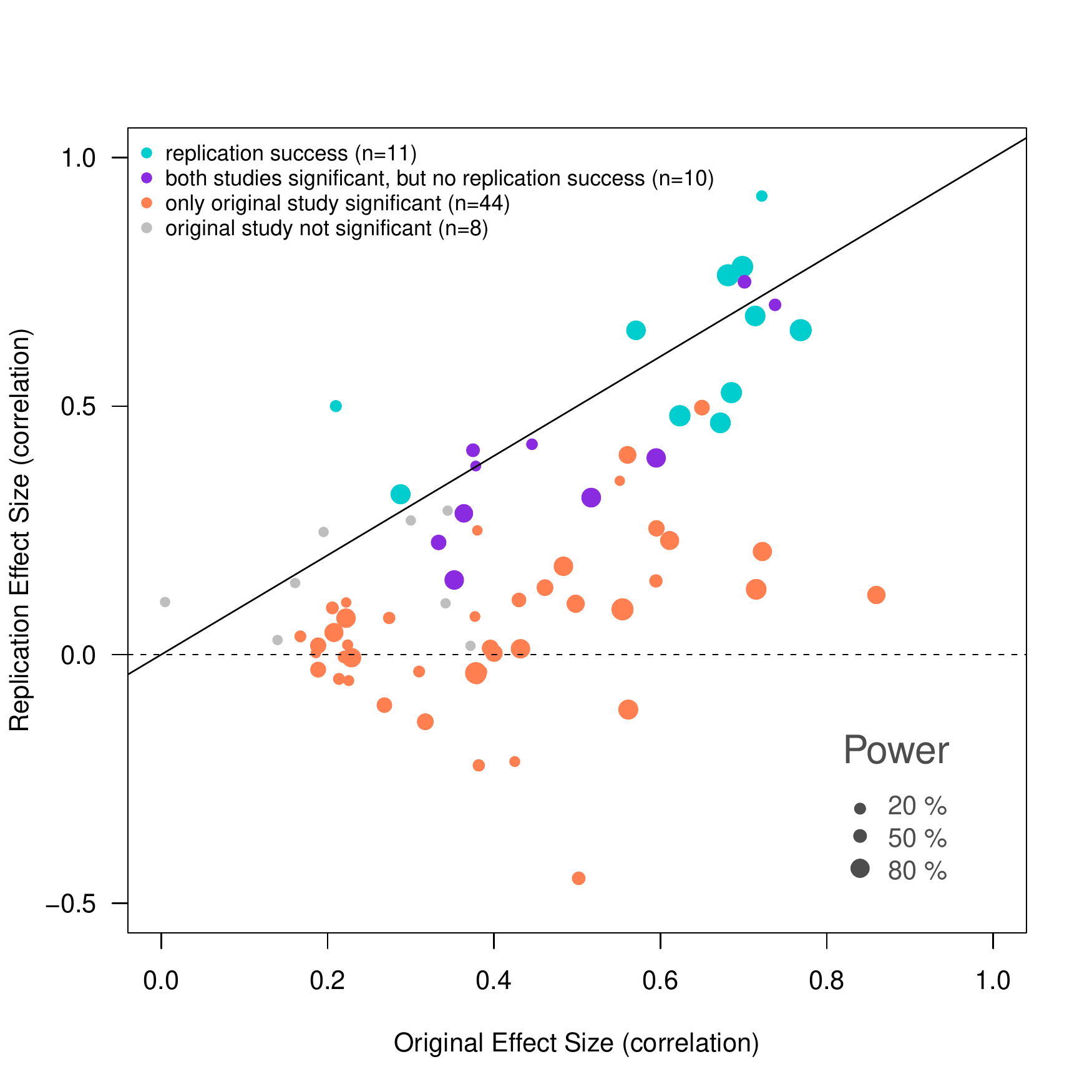} 

\end{knitrout}
\caption{Application to \citet{OSC2015}  data: The circles represent the effect
  estimates (correlations) of original and replication studies. The
  circle size represents the \hl{predictive} power for replication success at the two-sided 5\% level.  Replication success and significance is also assessed at
  the two-sided 5\% level and indicated by the color of the circles. \label{fig:fig8}}
\end{figure}

Figure \ref{fig:fig8} displays the replication versus the original
correlation estimates.  Eight of the 73
original studies are not significant at the standard
$\alpha=5\%$ level, three of them with $p$-values between 0.05 and
0.06. There have been 21 significant replication studies following from 
the 65 significant original studies. The sceptical $p$-value 
allows us to rank the studies by the degree of replication success. 
Table \ref{tab:tab1} lists the 24 most
successful replication studies with $p_S \leq 0.15$ of
which the top 11 have been successful
at the two-sided 5\% level ($p_S \leq 0.05$). The remaining 13 studies 
in Table \ref{tab:tab1} (with $p_S > 0.05$) show some interesting features. For example, 
study 18 has a non-significant replication result but 
still leads to replication success at the 10\% level. Conversely, there are several studies
with both $p_o \leq 0.05$ and $p_r \leq 0.05$ but $p_S > 0.10$.
This illustrates once again, that the sceptical $p$-value
does not only take significance of the original and replication study
into account, but also effect and sample sizes, both entering 
in the variance ratio $c$.

\begin{table}[ht]
\centering
\begin{tabular}{rrrrrrrrr}
   & \multicolumn{3}{c}{Original study}&
      \multicolumn{3}{c}{Replication study} & \multicolumn{2}{c}{Replication Success}  \\ \hline
    & $n_o$ & $\hat \rho_o$ & $p_o$ & $n_r$ & $\hat \rho_r$ & $p_r$ & $\mbox{Power}$  & $p_S$  \\ \hline
1 & 126 & 0.68 & < 0.0001 & 177 & 0.76 & < 0.0001 & > 99.9 & < 0.0001 \\ 
  2 &  78 & 0.77 & < 0.0001 &  38 & 0.65 & < 0.0001 & > 99.9 & < 0.0001 \\ 
  3 &  30 & 0.70 & < 0.0001 &  31 & 0.78 & < 0.0001 & 95.3 & 0.0005 \\ 
  4 & 174 & 0.29 & 0.0001 & 141 & 0.32 & < 0.0001 & 82.6 & 0.005 \\ 
  5 &  32 & 0.57 & 0.0005 &  32 & 0.65 & < 0.0001 & 78.7 & 0.007 \\ 
  6 &  22 & 0.71 & < 0.0001 &  22 & 0.68 & 0.0003 & 87.6 & 0.008 \\ 
  7 &  38 & 0.62 & < 0.0001 &  39 & 0.48 & 0.002 & 93.7 & 0.011 \\ 
  8 &  30 & 0.69 & < 0.0001 &  27 & 0.53 & 0.004 & 92.1 & 0.015 \\ 
  9 & 117 & 0.21 & 0.023 & 236 & 0.50 & < 0.0001 & 14.0 & 0.033 \\ 
  10 &  23 & 0.67 & 0.0003 &  31 & 0.47 & 0.007 & 88.4 & 0.038 \\ 
  11 &   9 & 0.72 & 0.026 &  18 & 0.92 & < 0.0001 & 9.3 & 0.048 \\ 
  12 & 154 & 0.36 & < 0.0001 &  50 & 0.28 & 0.045 & 69.7 & 0.052 \\ 
  13 &  40 & 0.37 & 0.017 &  95 & 0.41 & < 0.0001 & 29.0 & 0.06 \\ 
  14 &  11 & 0.70 & 0.014 &  11 & 0.75 & 0.006 & 28.9 & 0.067 \\ 
  15 &  25 & 0.59 & 0.001 &  33 & 0.40 & 0.022 & 76.7 & 0.072 \\ 
  16 &  41 & 0.52 & 0.0004 &  41 & 0.32 & 0.044 & 79.6 & 0.08 \\ 
  17 &   9 & 0.74 & 0.021 &  16 & 0.70 & 0.002 & 18.9 & 0.091 \\ 
  18 &  33 & 0.56 & 0.0005 &  21 & 0.40 & 0.071 & 64.5 & 0.096 \\ 
  19 &  33 & 0.38 & 0.029 &  72 & 0.38 & 0.0009 & 4.3 & 0.10 \\ 
  20 &  25 & 0.45 & 0.025 &  39 & 0.42 & 0.007 & 10.9 & 0.10 \\ 
  21 &  57 & 0.33 & 0.011 & 118 & 0.23 & 0.014 & 44.5 & 0.11 \\ 
  22 &  96 & 0.20 & 0.057 & 243 & 0.25 & < 0.0001 & 0.0 & 0.12 \\ 
  23 &  16 & 0.65 & 0.005 &  13 & 0.50 & 0.085 & 45.9 & 0.13 \\ 
  24 &  69 & 0.35 & 0.003 & 178 & 0.15 & 0.045 & 78.0 & 0.14 \\ 
   \hline
\end{tabular}
\caption{Results for the 24 most successful replication studies with $p_S \leq 0.15$, as listed in the last column. The penultimate column gives the predictive power  for replication success (in \%) at level $\alpha=5\%$.} 
\label{tab:tab1}
\end{table}

\section{Discussion}
\label{sec:sec6}
\hl{Science would proceed more efficiently if statistical approaches
  to inference are better aligned with scientific needs and practice
  \citep{Goodman2016b}. The traditional dichotomy between 'Bayesians'
  and 'frequentists' may not always be useful to achieve this goal.}  The
proposed approach follows the spirit of ``evolution rather than
revolution'' \citep{matthews2018} and provides a framework for
extracting more insight from replication studies based on standard
metrics (effect estimates, confidence intervals and $p$-values).
Instead of synthesising original and replication study results through
a meta-analysis, the original study result is challenged with the
sufficiently sceptical prior. Replication success is then defined as
conflict between the sufficiently sceptical prior and the replication
effect estimate. While the ordinary $p$-value quantifies the conflict
between the point null hypothesis and the replication data, the
sceptical $p$-value quantifies the conflict between the sufficiently
sceptical prior and the replication data.  It extends the ordinary
$p$-value of the replication study \hl{by taking into account effect
  and sample sizes of the two studies.}

\hl{ Just as the ordinary $p$-value is an indirect measure of the
  evidence against the null hypothesis, the sceptical $p$-value is an
  indirect measure of the degree of replication success. Specifically,
  a large sceptical $p$-value can occur if the replication sample size
  was too small, even if original and replication effect sizes are
  approximately equal, and should not be taken as evidence for no
  effect \citep{Altman485}.  This is not the only reason why it would
  be interesting to compare the sceptical $p$-value to direct
  ``forward-Bayes'' approaches, such as the replication Bayes Factor
  \citep{VerhagenWagenmakers2014,Ly_etal2018}, which quantifies the
  change in evidence brought about by observing the results from the
  replication study, given that the evidence from the original study
  is already available.}

Significance of both the original and the replication study is a
necessary but not sufficient requirement for replication success.  The
proposed framework thus \hl{extends}
the ``two pivotal study paradigm'' requiring two significant findings from
two independent confirmatory trials for regulatory drug approval, \hl{see
  \citet{kennedySchaffer:2017} for a recent review}. 
However, the difficulty to achieve replication success
if the evidence from the original study is only suggestive underlines
the need for more stringent $p$-value thresholds for claims of new
discoveries. The threshold for intrinsic credibility \eqref{eq:eq2} is
a natural choice for this task.


It would be interesting \hl{to extend the approach to a
  setting where several replication studies are available. For
  example, a summary estimate based on a meta-analysis of all
  available replication studies may be used to assess replication
  success. If results from replication studies become sequentially
  available, an alternative approach is} to combine original and
replication effect estimates into an overall summary measure.  Some
down-weighting of the original study result would in general be
required depending on the degree of replication success.  The overall
estimate could then be used as a new ``original'' effect estimate in
order to assess the success of a second replication study.  This would
open up new ways to iteratively challenge existing knowledge through a
series of replication studies and would provide an interesting
alternative to traditional evidence synthesis methods.

The proposed reverse-Bayes approach assumes a simple mathematical framework,
where likelihood, prior and posterior are all \hl{assumed to be normal}. It
will be of interest to extend this framework to other settings, for
example to the $t$-distribution. 

\paragraph*{Data and Software Availability} Data analyzed in this
article are originally from \citet{OSC2015} and have been downloaded
from \url{https://osf.io/fgjvw/}.  Software to compute the sceptical
$p$-value and the power or \hl{required} sample size \hl{to achieve replication success are}
available in the R-package \texttt{pCalibrate} available on the
Comprehensive R Archive Network
(\url{https://CRAN.R-project.org/package=pCalibrate}).

\paragraph*{Acknowledgments}
I am grateful to Robert Matthews, Ken Rice, Uri Simonsohn and the members of the
UZH Department of Biostatistics for helpful discussions and
suggestions. I also acknowledge helpful comments by referees and a
reviewer on a related grant proposal of mine.

\singlespacing
\bibliographystyle{apalike}
\bibliography{antritt}

\begin{thebibliography}{}

\bibitem[Altman and Bland, 1995]{Altman485}
Altman, D.~G. and Bland, J.~M. (1995).
\newblock Statistics notes: Absence of evidence is not evidence of absence.
\newblock {\em BMJ}, 311(7003):485.
\newblock \url{https://doi.org/10.1136/bmj.311.7003.485}.

\bibitem[Anderson and Maxwell, 2017]{Anderson2017}
Anderson, S.~F. and Maxwell, S.~E. (2017).
\newblock Addressing the {\textquotedblleft}replication
  crisis{\textquotedblright}: Using original studies to design replication
  studies with appropriate statistical power.
\newblock {\em Multivariate Behavioral Research}, 52(3):305--324.
\newblock \url{https://doi.org/10.1080/00273171.2017.1289361}.

\bibitem[Bayarri and Mayoral, 2002]{BayarriMayoral2002}
Bayarri, M.~J. and Mayoral, M. (2002).
\newblock Bayesian design of "successful" replications.
\newblock {\em The American Statistician}, 56:207--214.
\newblock \url{https://www.doi.org/10.1198/000313002155}.

\bibitem[Begley and Ioannidis, 2015]{Begley2015}
Begley, C.~G. and Ioannidis, J.~P. (2015).
\newblock Reproducibility in science.
\newblock {\em Circulation Research}, 116(1):116--126.
\newblock \url{https://circres.ahajournals.org/content/116/1/116}.

\bibitem[Benjamin et~al., 2018]{BenjaminEtAlinpress}
Benjamin, D.~J., Berger, J.~O., Johannesson, M., Nosek, B.~A., Wagenmakers,
  E.-J., et~al. (2018).
\newblock Redefine statistical significance.
\newblock {\em Nature Human Behaviour}, 2:6--10.
\newblock \url{https://dx.doi.org/10.1038/s41562-017-0189-z}.

\bibitem[Box, 1980]{box:1980}
Box, G. E.~P. (1980).
\newblock Sampling and {B}ayes' inference in scientific modelling and
  robustness (with discussion).
\newblock {\em Journal of the Royal Statistical Society, Series A},
  {143}:383--430.
\newblock \url{https://www.jstor.org/stable/2982063}.

\bibitem[Button et~al., 2013]{Button_etal2013}
Button, K.~S., Ioannidis, J. P.~A., Mokrysz, C., Nosek, B.~A., Flint, J.,
  Robinson, E. S.~J., and Munafò, M.~R. (2013).
\newblock Power failure: why small sample size undermines the reliability of
  neuroscience.
\newblock {\em Nature Reviews Neuroscience}, 14.
\newblock \url{https://dx.doi.org/10.1038/nrn3475}.

\bibitem[Camerer et~al., 2016]{Camerer2016}
Camerer, C.~F., Dreber, A., Forsell, E., Ho, T.-H., Huber, J., Johannesson, M.,
  Kirchler, M., Almenberg, J., Altmejd, A., Chan, T., Heikensten, E.,
  Holzmeister, F., Imai, T., Isaksson, S., Nave, G., Pfeiffer, T., Razen, M.,
  and Wu, H. (2016).
\newblock Evaluating replicability of laboratory experiments in economics.
\newblock {\em Science}, 351(6280):1433--1436.
\newblock \url{https://doi.org/10.1126/science.aaf0918}.

\bibitem[Camerer et~al., 2018]{Camerer2018}
Camerer, C.~F., Dreber, A., Holzmeister, F., Ho, T.-H., Huber, J., Johannesson,
  M., Kirchler, M., Nave, G., Nosek, B.~A., Pfeiffer, T., Altmejd, A.,
  Buttrick, N., Chan, T., Chen, Y., Forsell, E., Gampa, A., Heikensten, E.,
  Hummer, L., Imai, T., Isaksson, S., Manfredi, D., Rose, J., Wagenmakers,
  E.-J., and Wu, H. (2018).
\newblock {Evaluating the replicability of social science experiments in Nature
  and Science between 2010 and 2015}.
\newblock {\em Nature Human Behaviour}, 2(9):637--644.
\newblock \url{https://doi.org/10.1038/s41562-018-0399-z}.

\bibitem[Colquhoun, 2017]{Colquhoun:2017}
Colquhoun, D. (2017).
\newblock The reproducibility of research and the misinterpretation of
  p-values.
\newblock {\em Royal Society Open Science}, 4(12).
\newblock \url{https://dx.doi.org/10.1098/rsos.171085}.

\bibitem[Colquhoun, 2019]{Colquhoun:2019}
Colquhoun, D. (2019).
\newblock The false positive risk: A proposal concerning what to do about
  $p$-values.
\newblock {\em The American Statistician}, 73(sup1):192--201.
\newblock \url{https://doi.org/10.1080/00031305.2018.1529622}.

\bibitem[Evans and Moshonov, 2006]{evans.moshonov2006}
Evans, M. and Moshonov, H. (2006).
\newblock {C}hecking for prior-data conflict.
\newblock {\em Bayesian Analysis}, 1(4):893--914.
\newblock \url{https://projecteuclid.org/euclid.ba/1340370946}.

\bibitem[Fletcher et~al., 1993]{Fletcher1993}
Fletcher, A., Spiegelhalter, D., Staessen, J., Thijs, L., and Bulpitt, C.
  (1993).
\newblock Implications for trials in progress of publication of positive
  results.
\newblock {\em The Lancet}, 342(8872):653--657.
\newblock \url{https://doi.org/10.1016/0140-6736(93)91762-b}.

\bibitem[Good, 1950]{good:1950}
Good, I.~J. (1950).
\newblock {\em {Probability and the Weighing of Evidence}}.
\newblock Griffin, London, UK.

\bibitem[Good, 1983]{good:1983}
Good, I.~J. (1983).
\newblock {\em {Good Thinking: The Foundations of Probability and Its
  Applications}}.
\newblock University of Minnesota Press, Minneapolis.

\bibitem[Goodman, 1992]{Goodman1992}
Goodman, S.~N. (1992).
\newblock A comment on replication, p-values and evidence.
\newblock {\em Statistics in Medicine}, 11(7):875--879.
\newblock \url{https://doi.org/10.1002/sim.4780110705}.

\bibitem[Goodman, 2016]{Goodman2016b}
Goodman, S.~N. (2016).
\newblock Aligning statistical and scientific reasoning.
\newblock {\em Science}, 352(6290):1180--1181.

\bibitem[Goodman et~al., 2016]{Goodman2016}
Goodman, S.~N., Fanelli, D., and Ioannidis, J. P.~A. (2016).
\newblock What does research reproducibility mean?
\newblock {\em Science Translational Medicine}, 8(341):341ps12--341ps12.
\newblock \url{https://doi.org/10.1126/scitranslmed.aaf5027}.

\bibitem[Greenland, 2006]{greenland:2006}
Greenland, S. (2006).
\newblock Bayesian perspectives for epidemiological research: I. foundations
  and basic methods.
\newblock {\em International Journal of Epidemiology}, {35}:765--775.
\newblock \url{https://doi.org/10.1093/ije/dyi312}.

\bibitem[Greenland, 2011]{greenland:2011}
Greenland, S. (2011).
\newblock Null misinterpretation in statistical testing and its impact on
  health risk assessment.
\newblock {\em Preventive Medicine}, {53}:225--228.
\newblock \url{https://doi.org/10.1016/j.ypmed.2011.08.010}.

\bibitem[Held, 2013]{held2013}
Held, L. (2013).
\newblock {R}everse-{B}ayes analysis of two common misinterpretations of
  significance tests.
\newblock {\em Clinical Trials}, 10:236--242.
\newblock \url{https://doi.org/10.1177/1740774512468807}.

\bibitem[Held, 2019]{held2019}
Held, L. (2019).
\newblock The assessment of intrinsic credibility and a new argument for
  $p<0.005$.
\newblock {\em Royal Society Open Science}.
\newblock \url{https://doi.org/10.1098/rsos.181534}.

\bibitem[Ioannidis, 2005]{Ioannidis2005}
Ioannidis, J. P.~A. (2005).
\newblock Why most published research findings are false.
\newblock {\em {PLoS} Medicine}, 2(8):e124.
\newblock \url{https://doi.org/10.1371/journal.pmed.0020124}.

\bibitem[Ioannidis, 2018]{Ioannidis2018}
Ioannidis, J. P.~A. (2018).
\newblock The proposal to lower $p$-value thresholds to .005.
\newblock {\em JAMA: The Journal of the American Medical Association}.
\newblock \url{https://doi.org/10.1001/jama.2018.1536}.

\bibitem[Johnson, 2013]{Johnson2013}
Johnson, V.~E. (2013).
\newblock Revised standards for statistical evidence.
\newblock {\em Proceedings of the National Academy of Sciences of the United
  States of America}, 110(48):19313--19317.
\newblock \url{https://doi.org/10.1073/pnas.1313476110}.

\bibitem[Johnson et~al., 2016]{Johnson2016}
Johnson, V.~E., Payne, R.~D., Wang, T., Asher, A., and Mandal, S. (2016).
\newblock On the reproducibility of psychological science.
\newblock {\em Journal of the American Statistical Association},
  112(517):1--10.
\newblock \url{https://doi.org/10.1080/01621459.2016.1240079}.

\bibitem[Kennedy-Shaffer, 2017]{kennedySchaffer:2017}
Kennedy-Shaffer, L. (2017).
\newblock {When the Alpha is the Omega: $P$-values, "Substantial Evidence," and
  the 0.05 Standard at FDA}.
\newblock {\em Food and Drug Law Journal}, 72(4):595--635.
\newblock \url{https://www.ncbi.nlm.nih.gov/pmc/articles/PMC6169785/}.

\bibitem[Ly et~al., 2018]{Ly_etal2018}
Ly, A., Etz, A., Marsman, M., and Wagenmakers, E.-J. (2018).
\newblock Replication {B}ayes factors from evidence updating.
\newblock {\em Behavior Research Methods}.
\newblock \url{https://doi.org/10.3758/s13428-018-1092-x}.

\bibitem[Matthews, 2001a]{matthews:2001}
Matthews, R. A.~J. (2001a).
\newblock Methods for assessing the credibility of clinical trial outcomes.
\newblock {\em Drug Information Journal}, {35}:1469--1478.
\newblock \url{https://doi.org/10.1177/009286150103500442}.

\bibitem[Matthews, 2001b]{matthews:2001b}
Matthews, R. A.~J. (2001b).
\newblock Why {\em should} clinicians care about {B}ayesian methods? (with
  discussion).
\newblock {\em Journal of Statistical Planning and Inference}, {94}:43--71.
\newblock \url{https://doi.org/10.1016/S0378-3758(00)00232-9}.

\bibitem[Matthews, 2018]{matthews2018}
Matthews, R. A.~J. (2018).
\newblock Beyond 'significance': principles and practice of the analysis of
  credibility.
\newblock {\em Royal Society Open Science}, 5(1):171047.
\newblock \url{https://doi.org/10.1098/rsos.171047}.

\bibitem[O{\textquotesingle}Hagan et~al., 2005]{OHagan2005}
O{\textquotesingle}Hagan, A., Stevens, J.~W., and Campbell, M.~J. (2005).
\newblock Assurance in clinical trial design.
\newblock {\em Pharmaceutical Statistics}, 4(3):187--201.
\newblock \url{https://doi.org/10.1002/pst.175}.

\bibitem[{Open~Science~Collaboration}, 2015]{OSC2015}
{Open~Science~Collaboration} (2015).
\newblock Estimating the reproducibility of psychological science.
\newblock {\em Science}, 349:aac4716.
\newblock \url{https://doi.org/10.1126/science.aac4716}.

\bibitem[O’Hagan and Stevens, 2001]{OHaganStevens2001}
O’Hagan, A. and Stevens, J.~W. (2001).
\newblock Bayesian assessment of sample size for clinical trials of
  cost-effectiveness.
\newblock {\em Medical Decision Making}, 21(3):219--230.
\newblock \url{https://doi.org/10.1177/0272989X0102100307}.

\bibitem[Patil et~al., 2016]{Patil2016}
Patil, P., Peng, R.~D., and Leek, J.~T. (2016).
\newblock What should researchers expect when they replicate studies? {A}
  statistical view of replicability in psychological science.
\newblock {\em Perspectives on Psychological Science}, 11(4):539--544.
\newblock \url{https://doi.org/10.1177/1745691616646366}.

\bibitem[Sch\"{o}nbrodt and Wagenmakers, 2018]{Schoenbrodt2018}
Sch\"{o}nbrodt, F.~D. and Wagenmakers, E.-J. (2018).
\newblock Bayes factor design analysis: Planning for compelling evidence.
\newblock {\em Psychonomic Bulletin {\&} Review}, 25(1):128--142.
\newblock \url{https://doi.org/10.3758/s13423-017-1230-y}.

\bibitem[Simonsohn, 2015]{Simonsohn2015}
Simonsohn, U. (2015).
\newblock Small telescopes: Detectability and the evaluation of replication
  results.
\newblock {\em Psychological Science}, 26(5):559--569.
\newblock \url{https://doi.org/10.1177/0956797614567341}.

\bibitem[Spiegelhalter et~al., 2004]{sam:2004}
Spiegelhalter, D.~J., Abrams, K.~R., and Myles, J.~P. (2004).
\newblock {\em {Bayesian Approaches to Clinical Trials and Health-Care
  Evaluation}}.
\newblock Wiley, New York.

\bibitem[Spiegelhalter and Freedman, 1986]{SpiegelhalterFreedman1986}
Spiegelhalter, D.~J. and Freedman, L.~S. (1986).
\newblock A predictive approach to selecting the size of a clinical trial,
  based on subjective clinical opinion.
\newblock {\em Statistics in Medicine}, 5(1):1--13.
\newblock \url{https://doi.org/10.1002/sim.4780050103}.

\bibitem[Spiegelhalter et~al., 1986]{spiegelhalter1986}
Spiegelhalter, D.~J., Freedman, L.~S., and Blackburn, P.~R. (1986).
\newblock Monitoring clinical trials: conditional or predictive power?
\newblock {\em Controlled Clinical Trials}, 7(1):8--17.
\newblock \url{https://doi.org/10.1016/0197-2456(86)90003-6}.

\bibitem[Spiegelhalter et~al., 1994]{Spiegelhalter1994}
Spiegelhalter, D.~J., Freedman, L.~S., and Parmar, M. K.~B. (1994).
\newblock Bayesian approaches to randomized trials.
\newblock {\em Journal of the Royal Statistical Society. Series A (Statistics
  in Society)}, 157(3):357.
\newblock \url{https://doi.org/10.2307/2983527}.

\bibitem[van Aert and van Assen, 2017]{vanAertvanAssen2017}
van Aert, R. C.~M. and van Assen, M. A. L.~M. (2017).
\newblock Bayesian evaluation of effect size after replicating an original
  study.
\newblock {\em PLOS ONE}, 12(4):1--23.
\newblock \url{https://doi.org/10.1371/journal.pone.0175302}.

\bibitem[Verhagen and Wagenmakers, 2014]{VerhagenWagenmakers2014}
Verhagen, J. and Wagenmakers, E.-J. (2014).
\newblock Bayesian tests to quantify the result of a replication attempt.
\newblock {\em Journal of Experimental Psychology}, {143}:1457--1475.
\newblock \url{https://dx.doi.org/10.1037/a0036731}.

\end{thebibliography}

\appendix{}
\onehalfspacing
\part*{\appendixname}
\pdfbookmark[0]{\appendixname}{\appendixname}
\section{Proof of equation \eqref{eq:extrinsic.p}}\label{app:app1}  
We have
\[
    t_{\mbox{\scriptsize Box}}^2 = \frac{\hat \theta_r^2}{{\tau^2+\sigma_r^2}} 
  = \frac{\hat \theta_r^2}{\sigma_r^2} \left(\frac{c}{t_o^2/z_{\alpha/2}^2 - 1} + 1 \right)^{-1} 
  = t_r^2 \left(\frac{t_o^2/z_{\alpha/2}^2 - 1}{c + t_o^2/z_{\alpha/2}^2 - 1} \right), 
\]
so the requirement $t_{\mbox{\scriptsize Box}}^2 \geq z_{\alpha/2}^2$ for replication success \hl{at level $\alpha$} is equivalent to 
\[
\frac{t_r^2}{z_{\alpha/2}^2} \left(\frac{t_o^2}{z_{\alpha/2}^2}-1 \right) \geq c + t_o^2/z_{\alpha/2}^2 - 1.
\]
Subtracting $t_o^2/z_{\alpha/2}^2 - 1$ on both sides leads to 
\eqref{eq:extrinsic.p}.

\section{The limiting cases $\sigma_o^2 \downarrow 0$ and $\sigma_r^2 \downarrow 0$}
\label{app:app2}  
Equation \eqref{eq:quadEq} can be re-written as 
\begin{equation}\label{eq:help}
\frac{c-1}{t_A^2} z_S^4 + 2 \, z_S^2 =  t^2_H ,
\end{equation}
where 
$$\frac{c-1}{t_A^2} = \frac{\sigma_o^2 - \sigma_r^2}{\sigma_r^2} \frac{2}{t_o^2+t_r^2} = 
\frac{2 \, \sigma_o^2 (\sigma_o^2 - \sigma_r^2)}{\hat \theta_o^2 \sigma_r^2+\hat \theta_r^2 \sigma_o^2}.$$ 
For $\sigma_o^2 \downarrow 0$ we thus have ${(c-1)}/{t_A^2} \rightarrow 0$
and 
$t^2_H \rightarrow 2 \, t_r^2$ so equation \eqref{eq:help} reduces to
$2 \, z_S^2 =  2 \, t_r^2$ and hence $z_S^2 =  t_r^2$.
For $\sigma_r^2 \downarrow 0$ we have 
${(c-1)}/{t_A^2} \rightarrow {(2 \, \sigma_o^2)}/{\hat \theta_r^2}$
and $t^2_H \rightarrow 2 \, t_o^2$ so equation \eqref{eq:help} reduces to
$({\sigma_o^2}/{\hat \theta_r^2}) \, z_S^4 + z_S^2 =  t^2_o$.
The solution of this equation is 
\begin{eqnarray*}
  z_S^2 & = & \frac{\hat \theta_r^2}{2 \, \sigma_o^2} \left\{ \sqrt{1 + 4 \sigma_o^2 t_o^2/\hat \theta_r^2} -1 \right\} \\
& = & \frac{\hat \theta_r^2}{2 \, \sigma_o^2} \left\{ \sqrt{1 + 4 / d} -1 \right\} \\
& = & \frac{\hat \theta_o^2}{2 \, \sigma_o^2} \left\{ \sqrt{d^2 + 4 \, d} - d \right\} \\
& = & \frac{t_o^2}{2} \left\{ \sqrt{d(d + 4)} - d \right\}, 
\end{eqnarray*}
which is equation \eqref{eq:p.matthews} with $d= \hat \theta_r^2/\hat \theta_o^2$.

\section{Proof of result \eqref{eq:eq1}}\label{app:app3}  
Equation \eqref{eq:solution} reduces for $c \downarrow 0$ to 
\[
z_S^2 =  {t_A^2} - \sqrt{t_A^2 \left[{t_A^2} - t^2_H\right]} 
=  {t_A^2} - \sqrt{\frac{t_A^2 }{2}\frac{(t_o^2 - t_r^2)^2}{t_o^2 + t_r^2}} 
= {t_A^2} - \frac{\abs{t_o^2-t_r^2}}{2} 
= \min\{t_o^2, t_r^2\}.
\]

The derivative of \eqref{eq:solution} with respect to $c$ is (for $c \neq 1$)
\begin{eqnarray}
\frac{d \, z_S^2}{d \, c} &=& - \frac{1}{c-1}
\left\{ z_S^2 - \frac{1}{2}\frac{t_A^2 t_H^2}{(c-1) z_S^2 + t_A^2}
\right\} \nonumber \\
&=& - \frac{z_S^2}{c-1}
\left\{ 1 - \frac{1}{2}\frac{(c-1) z_S^2 + 2 \, t_A^2}{(c-1) z_S^2 + t_A^2}
\right\} \nonumber \\
&=& - \frac{1}{2} \, \frac{z_S^4}{(c-1) z_S^2 + t_A^2} \label{eq:derivative}
\end{eqnarray}
where the middle line follows from \eqref{eq:quadEq} and the last line also holds for $c=1$.  It is easy to
see from \eqref{eq:solution} that $(c-1) z_S ^2 + t_A^2 > 0$ for all $c$,
and therefore \eqref{eq:derivative} is negative for all $c$.

\section{Proof of results in Section \ref{sec:design}}\label{app:app4}  

For notational simplicity I omit the conditioning on $\hat
\theta_o$ in the following.  
Equation \eqref{eq:priorPredictive} implies a distribution on $t_r =
\hat \theta_r / \sigma_r = \sqrt{n_r} \, \hat \theta_r / \kappa$,
\[
t_r  \sim \Nor\left(
\sqrt{n_r} \, \frac{\hat \theta_o}{\kappa}, \frac{n_o + n_r}{n_o} \right),
\]
so $t_r = \sqrt{{(n_o+n_r)}/{n_o}} \, \tilde t_r$
where
$$
 \tilde t_r \sim \Nor\left(\sqrt{\frac{n_o n_r}{n_o + n_r}} \, \frac{\hat \theta_o}{\kappa}, 1 \right).
$$
Therefore $t_r^2 = {{(n_o+n_r)}/{n_o}} \cdot \tilde t_r^2$ follows a scaled non-central $\chi^2$-distribution with 1 degree of freedom, 
scaling factor ${{(n_o+n_r)}/{n_o}}$ $\hl{=1+c}$ and non-centrality parameter 
$
\lambda = {(n_o n_r)}/{(n_o + n_r)} \cdot {\hat \theta_o^2}/{\kappa^2}$ $\hl{ = t_o^2/(1+1/c)}.
$

Things simplify somewhat for a point prior $\theta=\hat \theta_o$ at the 
estimate from the original study. Then 
$  \hat \theta_r \given \hat \theta_o \sim \Nor\left(\hat
\theta_o, {\kappa^2}/{n_r} \right)$
so
$t_r  \sim \Nor\left(
\sqrt{n_r} \, {\hat \theta_o}/{\kappa}, 1 \right)$.
Now $t_r^2$ follows a non-central $\chi^2$-distribution with 1 degree of freedom and non-centrality parameter 
$\lambda = n_r \, {\hat \theta_o^2}/{\kappa^2}$ $\hl{ = c \, t_o^2}$.

\end{document}